\documentclass{sig-alternate}
\newfont{\mycrnotice}{ptmr8t at 7pt}
\newfont{\myconfname}{ptmri8t at 7pt}
\let\crnotice\mycrnotice%
\let\confname\myconfname%

\usepackage[german,american]{babel}
\usepackage[T1]{fontenc}
\usepackage[utf8]{inputenc}
\usepackage{times}
\usepackage{textcomp}
\usepackage{graphicx}
\usepackage{amssymb}
\usepackage{microtype} 

\usepackage{amsmath,stackengine}
\usepackage{multirow}

\usepackage{array}
\newcolumntype{L}[1]{>{\centering}m{#1}}

\usepackage{url}
\renewcommand{\tt}{\fontencoding{OT1}\fontfamily{cmtt}\selectfont}

\usepackage[usenames,dvipsnames]{color}
\usepackage{booktabs}
\usepackage[numbers,sectionbib]{natbib}
\makeatletter
\def\@copyrightspace{\relax}
\makeatother

\stackMath
\begin{document}

\title{Detecting Clickbait in Online Social Media: You Won't Believe How We Did It}
\subtitle{The Salmon Clickbait Detector at the Clickbait Challenge 2017}

\numberofauthors{1}
\author{
\alignauthor
Aviad Elyashar, Jorge Bendahan, and Rami Puzis\\
\affaddr{Telekom Innovation Laboratories\\ 
	Department of Software and Information Systems Engineering\\
	Ben-Gurion University of the Negev, Beer-Sheva, Israel}\\
\affaddr{aviade@post.bgu.ac.il, jorgeaug@post.bgu.ac.il, puzis@bgu.ac.il}\\
\alignauthor
}

\maketitle

\begin{abstract}
In this paper, we propose an approach for the detection of \emph{clickbait posts} in online social media (OSM). 
\emph{Clickbait posts} are short catchy phrases that attract a user's attention to click to an article.  
The approach is based on a machine learning (ML) classifier capable of distinguishing between \emph{clickbait} and \emph{legitimate posts} published in OSM.
The suggested classifier is based on a variety of features, including image related features, linguistic analysis, and methods for \emph{abuser} detection. 
In order to evaluate our method, we used two datasets provided by Clickbait Challenge 2017.
The best performance obtained by the ML classifier was an AUC of 0.8, accuracy of 0.812, precision of 0.819, and recall of 0.966.
In addition, as opposed to previous studies, we found that \emph{clickbait post} titles are statistically significant shorter than \emph{legitimate post} titles. 
Finally, we found that counting the number of formal English words in the given content
is useful for clickbait detection. 

\end{abstract}

\section{Introduction}
\label{sec:introduction}

In the past, offline media outlets, such as newspapers, were the main source of information used to inform people.
However, in recent decades, these traditional news outlets have been replaced with online resources.
We can attribute this change to the great diversity of options ranging from local, national, and international online media outlets to several niche blogs, which focus on a specific area of interest offered~\cite{chakraborty2016stop}, to the large numbers of readers using smart devices, to content generators, which provide users personalized news, derived from a wide variety of news sources~\cite{lavie2010user}, and to the fact that most online media websites do not charge a fee for their services, as opposed to traditional media outlets~\cite{chakraborty2016stop}.
As a result, online news is rapidly replacing traditional media outlets~\cite{lavie2010user}. 

Although online news provides numerous benefits, this domain also has problematic issues.
In most cases, the revenues of online media websites are not based on subscribers' charges, but instead are based on advertisements, which are published on the websites~\cite{chakraborty2016stop}.  
These  results in a significant amount of competition among the online media outlets that vie for readers' attention and their clicks which increase the online media websites' income.
Therefore, in order to attract users and encourage them to visit the online media website and click on a given article, the website administrators use a variety of techniques, including the use of catchy headlines along with the article links, which lure users in to clicking on the links~\cite{chakraborty2016stop}.

Such short teaser posts are known as clickbait, and this type of Web content is effectively an advertisement aimed attracting visitors' attention and encouraging them to click on an attached link, which directs the user to a specific Web page~\cite{potthast2016clickbait}. 
Clickbait is commonly spread via online social media (OSM)~\cite{chen2015misleading}.
In recent years, the use of clickbait has contributed to, and exacerbated, the rapid spread of rumors and misinformation online~\cite{chen2015misleading}. 
In many cases, clickbait is characterized by an anticipated emotional reaction, and lack of knowledge (e.g., "15 surprising facts about Tesla cars you probably didn't know", and "Here's what people really thought about that Trump press conference").

Clickbait takes advantage of the cognitive phenomenon known as the \emph{Curiosity Gap}~\cite{loewenstein1994psychology}.
Clickbait headlines provide referencing cues, which create curiosity among users. 
This curiosity encourages the readers to click on the link in order to address the knowledge gap~\cite{chakraborty2016stop}.  

Currently, the state of the art solutions for automatic clickbait detection are based on machine learning (ML) techniques, yet many of these studies suffer low accuracy. 
In this study, we propose a method for detecting \emph{clickbait posts} in OSM based on a ML classifier capable of distinguishing between \emph{clickbait} and \emph{legitimate posts} published in OSM.   
The classifier is based on a variety of features, including image related features, linguistic analysis, and methods for \emph{abuser} detection. 
In order to evaluate our method, we used two datasets provided by Clickbait Challenge 2017~\cite{potthast:2017a}.   

The contributions of this paper are as followes:
\begin{itemize}
	
	\item We identified useful new features, which combine the information extracted from both the post, and its article. 
	These features were found among the top essential features in both datasets for detecting \emph{clickbait} in OSM.  
	To the best of our knowledge, we are the first to suggest features that use both components together;
	\item As opposed to previous studies, which concluded that benign content is shorter than malicious, we found that \emph{clickbait post} titles are statistically significant shorter than \emph{legitimate post} titles in the two given datasets. 
	\item We found that counting the number of formal English words in given content is useful feature for clickbait detection. 
	To the best of our knowledge this feature is new;
	\item We presented evidence that the post's title is the most important component to use for detecting clickbait.
\end{itemize}

The rest of this paper is organized as follows: 
In Section~\ref{sec:related_work} we review well-known methods for the detection of clickbait, as well as \emph{abusers}, in OSM.  
We explain our approach for detecting clickbait in OSM in Section~\ref{sec:approach}.
We present the results of the evaluation carried out on the datasets in Section~\ref{sec:evaluation_results}.  
Finally, we conclude the paper in Section~\ref{sec:conclusion}. 

\section{Related Work}
\label{sec:related_work}

In this section, we provide background information regarding the major issues focused on this study: clickbait detection methods, and methods for identifying \emph{abusers} in OSM. 

\subsection{Clickbait Detection}
\label{sec:clickbait_detection}

In 2014, Vijgen~\cite{vijgen2014listicle} studied articles that collect lists of things, called "listicles".
In many cases, these listicles are suspected to be clickbait due to their titles, which are typically shared as teaser messages.
Vijgen collected 720 listicles published by BuzzFeed~\cite{buzzfeed} in January 2014.  
He found that all of the titles contain a cardinal number, which is the same as the number of items listed.
In addition, the titles contained strong nouns and adjectives that convey authority and sensationalism.

In the same year, Gianotto~\cite{gianotto2014downworthy} implemented a browser plugin that detect clickbait based on a rule set.

In 2015, Blom and Hansen~\cite{blom2015click} mapped the use of forward-referencing headlines in online news by analyzing 100,000 headlines published in ten different Danish news websites.
They found that commercialization and tabloidization seem to lead to the recurrent use of forward-referencing in Danish online news headlines.

Also in 2015, Chen et al.~\cite{chen2015misleading} examined optional methods for the automatic detection of clickbait.
They divided the methods in to methods that rely on content cues and those that use non-text cues. 
The former includes lexical and semantic analysis, as well as syntactic analysis, whereas the latter includes image and user behavior analysis. 
They suggested that a hybrid approach which merges both methods may yield better results.  
 
In 2016, Potthast et al.~\cite{potthast2016clickbait} proposed a model for detecting clickbait automatically.
They collected and annotated a corpus of 2,992 tweets and developed a ML classifier, which attempts to detect clickbait. 
Their model was based on 215 features, including image, sentiment, and linguistic analysis, as well as extracting Twitter-specific features and bag-of-words features.
They achieved results: 0.79 AUC, 0.76 precision, and recall of 0.76 with a Random Forest classifier.
The researchers analyzed also the web pages linked from a given tweet. 
Their analysis included measurement of the main content word length.
Similarly, we extracted from the targeted articles features (e.g., word length), however, we innovated new features that combines the information from both the post and the article.

In the same year, Chakraborty et al.~\cite{chakraborty2016stop} created a ML classifier for detecting clickbait automatically and implemented a browser extension called `Stop Clickbait' to prevent readers from reading clickbait.
For training a classifier, they extracted the headlines from a corpus of 18,513 Wikinews articles as \emph{legitimate posts}, and for clickbait they crawled 8,069 Web articles from several Web domains, such as BuzzFeed~\cite{buzzfeed}, ViralNova~\cite{viralnova}, ScoopwHoop~\cite{scoopwhoop}, and ViralStories~\cite{viralstories}. 
 They used a set of fourteen features spanning linguistic analysis, word patterns, and N-gram to train their classifier.  
They reported 93\% and, 89\% and accuracy in detecting and blocking clickbait (respectively) by using a Support Vector Machine (SVM) classifier.

More recently, in 2017, Chakraborty et al.~\cite{chakraborty2017tabloids} analyzed the social sharing patterns of \emph{clickbait} and \emph{legitimate posts} in Twitter by collecting a dataset from Twitter. 

\subsection{Abuser Detection}
\label{sec:abuser_detection}
There is great similarity between detecting \emph{abusers}, and \emph{clickbait posts} in OSM.
The posts of \emph{abusers} and \emph{clickbait} attempt to manipulate the environment.
For example, \emph{abusers} in Twitter manipulate account popularity with artificial retweets~\cite{song2015crowdtarget}.
In a similar way, online media websites use \emph{clickbait posts} in order to attract readers' attention by publishing posts that contain misleading titles that exaggerate the content of the targeted article or try to for look like credible journalism~\cite{tan2017clickbait}.

Several studies involving the identification of abusers have been conducted. 
In 2012, Cao et al.~\cite{cao2012aiding} proposed a method that clusters users according to the similarity of the posted URL and then classifies each cluster as either malicious or not by extracting clusters' behavioral and content features.

In 2013, a method for the identification of crowdturfers on Twitter was presented by Lee at al.~\cite{lee2013crowdturfers}.  
They extracted features that were related to account properties, activity patterns, and linguistic properties.

In 2015, Song et al.~\cite{song2015crowdtarget} proposed a new approach for the detection of artificially promoted objects, such as posts, pages, and hyper-links.
They analyzed the characteristics of the target instead of its accounts and created three classifiers, achieving a true positive rate of 0.98. 

Dickerson et al.~\cite{dickerson2014using} and Davis et al.~\cite{davis2016botornot} used ML techniques for bot detection. 
\cite{dickerson2014using} based their detection on sentiment analysis, social network analysis, posted content, and account property features.
\cite{davis2016botornot} presented BotOrNot, a bot identification platform that can be used through a Web user interface. 
They detected bots based on all of the features used by~\cite{dickerson2014using}, as well as behavior features.

Recently, Tacchini et al.~\cite{tacchini2017some} identified hoaxes within Facebook based on the users who interacted with these hoaxes rather than the hoaxes' content.
Elyashar et al.~\cite{elyashar2017measurement} proposed a method for estimating the authenticity of online discussions based on several similarity functions of OSM accounts participating in an online discussion.
They found that the similarity function with the best performance across all of the datasets was the bag-of-words.


\section{Approach}
\label{sec:approach}

We propose an approach for differentiating between \emph{clickbait} and \emph{legitimate posts} based on image to text features, linguistic and behavior analysis.   
In this section, we provide a comprehensive description of the given datasets, as well as the proposed method. 

\subsection{Datasets}
\label{sec:datasets}

Two training datasets were used for evaluation of our proposed method.
The first dataset is defined as a small initial training set~\cite{potthast:2016}. 
In total, it includes 2,495 posts, among them 762 \emph{clickbait posts}, and 1,697 \emph{legitimate posts}.  

The second dataset is defined as a large training and validation dataset~\cite{potthast:2017b}.
It includes a total of 19,538 posts, among them 4,761 \emph{clickbait posts}, and 14,777 \emph{legitimate posts}.

\begin{figure}[htb]
	\centering
	\includegraphics[width=1\linewidth]{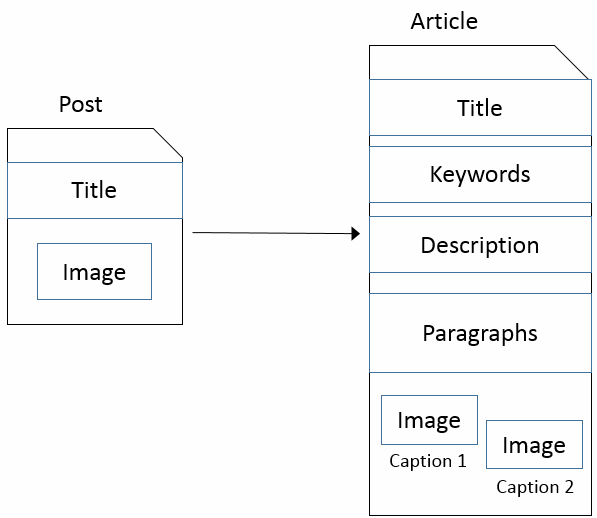}
	\caption{The structure of the records in the datasets provided}
	\label{fig:problem_structure}	
\end{figure}

Each post may include an attached image, and must point to a targeted article.
The targeted article includes a title, description, paragraphs, and captions attached to the images.
Figure~\ref{fig:problem_structure} presents the structure of the instances in the datasets.

\subsection{Content Related Features}
\label{sec:content_related_features}

We begin this section with a few definitions.  
Let \(P\) denote the collection of posts published in OSM; \(p\) is defined as a post in post collection \(P\).
For every post \(p\in P\), \(article(p)\) denotes the targeted article, which post \(p\) points to.
We define the following functions:

1) \(img(p)\) is defined as a function of \(P\). 
This function maps every post to its image if it presents; otherwise it is null.
 
2) \(OCR(img(p))\) is defined as a function of the image of post \(p\in P\).
It maps every post to the text extracted from the image, if an image is present; otherwise it is null.
 
3) \(title(p)\) or \(title(article(p))\) is defined as a function of post \(p\in P\) and the article which post \(p\) points to.
It maps every post and post's article to the text extracted from the post's title and article's title respectively.

4) \(description(article(p))\) is a function of the article which post \(p\) points to.
It maps every article's post to the text extracted from the article's description.

5) \(keywords(article(p))\) is a function of the article which is post \(p\) points to.
It maps every article to its predefined keywords.

6) \(paragraphs(article(p))\) is a function of the article which post \(p\) points to .
It maps every article to the text extracted from the article's paragraphs.

7) \(captions(article(p))\) is a function of the article which post \(p\) points to .
It maps every article to the text extracted from the article's captions.

8) \(len_{characters}(content)\) is a function of the given content.
The content can be \(title(p)\), \(title(article(p))\), \(description(article(p))\), \(keywords(article(p))\), \(paragraphs(article(p))\), or \(captions(article(p))\).
It returns the number of characters in given content (between 0 to infinity).
If the content is plural like keywords, paragraphs, and captions, it returns the average number of characters. 
When there is no available content the function returns -1. 

9) \(len_{words}(content)\) is a function of the same content described above.
The only difference is that this function returns the number of words in given content (between 0 to infinity). 
When there is no available content the function returns -1. 

10) \(words(content)\) is a function on the given content.
The content can be the same as described above in definition 8.
It returns a set which is a well-defined collection of distinct objects. 
In our case, the set is made up of words which comprise the given content.

11) \(lang\text{-}dict_{formal}(words(content))\) is a function on a set of words.
It returns a set of the formal English words from the given words.

12) \(lang\text{-}dict_{informal}(words(content))\) is a function on a set of words.
It returns a set of informal English words from the given words.

\subsubsection{Image Related Features}
\label{sec:image_related_features}
Images are usually processed before reading the full article or post and can be used to attract readers’ attention~\cite{ecker2014effects}.
One of the methods used to effectively manipulate emotion in readers is based on the use of images through the proximity of the images in the post, and more specifically, the proximity of the images to the headline of post~\cite{chen2015misleading}.  
In order to extract text from the post's image, we used pytesseract (Python-tesseract) package~\cite{pytesseract}, an optical character recognition (OCR) tool available in Python aimed at recognizing the text embedded in images.


\paragraph{Presence of an Image in a Post}
\emph{Clickbait posts} in Twitter were found to contain a significantly greater proportion of images than \emph{legitimate posts}~\cite{chakraborty2017tabloids}.
Therefore, we think it would be interesting to evaluate whether the presence of an image in a post is a useful feature for detecting clickbait.

	\[
	image\text{-}presence(x)= 
	\begin{cases}
	1,& \text{if } img(p) \neq null \\
	0,              & \text{otherwise}
	\end{cases}
	\]

\paragraph{Presence of Text in a Post's Image}
This feature is related to the text that may appear in a post's image.
We would like to understand whether the images in \emph{clickbait posts} contains more text than the posts of \emph{non-clickbait}.

 	\[
 	text\text{-}in\text{-}image(x)= 
 	\begin{cases}
 	1,& \text{if } img(p) \neq null\quad \& \quad OCR(img(p)) \neq null \\
 	0,              & \text{otherwise}
 	\end{cases}
 	\]

%

\subsubsection{Linguistic Analysis}
\label{sec:linguistic_analysis}
Linguistic analysis is another well-known method for detecting \emph{clickbait}.
This type of analysis includes semantic and syntactic analysis in order to find nuances that occur more frequently in \emph{clickbait posts} compared to \emph{legitimate posts}~\cite{chakraborty2016stop}.

\paragraph{Number of Characters}
\label{sec:number_of_characters}
\emph{Non-clickbait post} titles were found to be shorter than \emph{clickbait} titles~\cite{chakraborty2016stop}. 
Therefore, we count the number of characters to each content item.

\begin{equation}
num\text{-}of\text{-}characters(x) = \stackunder[5pt]{}{%
	\begin{cases}
	len_{characters}(x),& \text{if } \quad \exists x \\
	-1,              & \text{otherwise}
	\end{cases}}
\end{equation}

From this function we extracted the following features: the \emph{number of characters in post's title}, the \emph{number of characters in text extracted from post's image}, the \emph{number of characters in article's title}, the \emph{number of characters in article's description}, the \emph{number of characters in article's keywords}, the \emph{number of characters in article's captions}, and the \emph{number of characters in article's paragraphs}.

\paragraph{Difference Between Number of Characters} 
\label{sec:diff_number_of_characters}
This function measures the difference between the number of characters in two content elements.
As opposed to previous studies, which simply measured the length of the suspected post's title, with this feature we measure the relation between each pair of content elements. 
It would be interesting to understand whether there is a connection between the number of characters of a content field from the post and the targeted article.

	\begin{equation}
	\begin{aligned}
	diff\text{-}num\text{-}of\text{-}characters(cont_x,cont_y) ={} &  \\
	|num\text{-}of\text{-}characters(cont_x)-num\text{-}of\text{-}characters(cont_y)|&  
	\end{aligned}
	\end{equation}

Based on this function we can calculate features between each pair of content items.
There is no difference whether we calculated the difference number of characters between \(cont_x\), and \(cont_y\), or the difference number of characters between \(cont_y\), and \(cont_x\) due to the absolute value, therefore 
\begin{eqnarray*}
	diff\text{-}num\text{-}of\text{-}characters(cont_x,cont_y) = \\ 
	diff\text{-}num\text{-}of\text{-}characters(cont_y, cont_x) 
\end{eqnarray*} 

Thus, the number of combinations is a triangular number when n=6.
In total, we extracted twenty-one features, including: the \emph{diff number of characters between post's title and article's title},
the \emph{diff num of characters between post's title and article's description}, the \emph{diff num of characters between post's title and article's keywords},..., and the \emph{diff num of characters between article's paragraphs and text extracted from post's image}. 

\paragraph{Number of Characters Ratio} 
\label{sec:ratio_number_of_characters}
One more function we can extract from the basic \emph{number of characters} function is the ratio between the number of characters of two content elements.

\begin{equation}
\def\stackalignment{l}
\stackunder[5pt]{num\text{-}of\text{-}charcters\text{-}ratio(cont_x,cont_y) =}{
	\begin{cases}
	\mathopen|\frac{num\text{-}of\text{-}charcters(cont_x)}{num\text{-}of\text{-}characters(cont_y)}\mathclose|, & \text{if} \quad \exists cont_x \quad \& \quad \exists cont_y\\
	-1, & \text{otherwise}
	\end{cases}}
\end{equation}
 
We calculate only one side of the equation.
The other side is the opposite of the first.
This means that
\begin{eqnarray*}
	num\text{-}of\text{-}characters\text{-}ratio(cont_x,cont_y) = \\ 
	\frac{1}{num\text{-}of\text{-}characters\text{-}ratio(cont_y, cont_x)}
\end{eqnarray*}

Thus, the number of combinations is a triangular number when n=6.
In total, we extracted twenty-one features, including: \emph{number of characters ratio between post's title and article's title},
\emph{number of characters ratio between post's title and article's description}, \emph{number of characters ratio between post's title and article's keywords},..., \emph{number of characters ratio between article's paragraphs and text extracted from post's image}. 

\paragraph{Number of Words}
\label{sec:number_of_words}

Chakraborty et al.~\cite{chakraborty2016stop} found that there are more words in \emph{clickbait} titles than \emph{non-clickbait} titles.
This function count the number of words for each content \(x\). 
If the content does not exist then it returns -1.

\begin{equation}
num\text{-}of\text{-}words(x) = \stackunder[5pt]{}{%
	\begin{cases}
	len_{words}(x),& \text{if } \quad \exists x \\
	-1,              & \text{otherwise}
	\end{cases}}
\end{equation}

Based on this function we extracted the following features: the \emph{number of words in post's title}, the \emph{number of words in text extracted from post's image}, the \emph{number of words in article's title}, th \emph{number of words in article's description}, the \emph{number of words in article's keywords}, and the\emph{number of words in article's captions, and number of words in article's paragraphs}.

\paragraph{Difference Between Number of Words} 
\label{sec:diff_number_of_words}
Given two content elements, this function determines the difference between the number of words in each element.

\begin{equation}
\begin{aligned}
diff\text{-}num\text{-}of\text{-}words(cont_x,cont_y) ={} &  \\
|num\text{-}of\text{-}words(cont_x)-num\text{-}of\text{-}characters(cont_y)|&  
\end{aligned}
\end{equation}

This function is similar to the function described in Section~\ref{sec:diff_number_of_characters}.
The only difference is that this function focuses on words, while in Section~\ref{sec:diff_number_of_words} the function focuses on characters.

\paragraph{Number of Words Ratio} 
\label{sec:ratio_number_of_words}
Another function we can extract from the basic function described in Section~\ref{sec:diff_number_of_words}  is the ratio between the number of words in two content elements. 

\begin{equation}
\def\stackalignment{l}
\stackunder[5pt]{num\text{-}of\text{-}words\text{-}ratio(cont_x,cont_y) =}{
	\begin{cases}
	\mathopen|\frac{num\text{-}of\text{-}words(cont_x)}{num\text{-}of\text{-}words(cont_y)}\mathclose|, & \text{if} \quad \exists cont_x \quad \& \quad \exists cont_y\\
	-1, & \text{otherwise}
	\end{cases}}
\end{equation}

We used this function the same way we used the similar function presented in Section~\ref{sec:ratio_number_of_characters}.

\paragraph{Common Words Between Article Keywords and Others} 
\label{sec:common_words_between_article_keywords_and_others}
In many cases, \emph{clickbait posts} were found to contain misleading titles, exaggerating the content of the targeted article~\cite{biyani20168}. 
We would like to look at this more closely to detect the nuances of this exaggeration.
One of the methods that we used focuses on an article's keywords.  
An article's keywords should reflect the main issues of a given article.
The suggested function measures how often the article's keywords exist in other content elements (e.g., the post's title).   

\begin{equation}
\begin{aligned}
num\text{-}of\text{-}common\text{-}article\text{-}words(cont_x) ={} &  \\
keywords(article(p)) \cap words(cont_x) &  
\end{aligned}
\end{equation}

Based on this function we extracted the following features: the \emph{number of common words between article keywords and post's title}, the \emph{number of common words between article keywords and text extracted from post's image}, the \emph{number of common words between article keywords and article keywords}, the \emph{number of common words between article keywords and article description}, the \emph{number of common words between article keywords and article captions}, the \emph{number of common words between article keywords and article paragraphs}. 

\paragraph{Number of Formal and Informal English Words} 
\label{sec:number_of_formal_and_informal_words}
In the advertisement environment, slang or profane words are commonly used to get users' attention~\cite{zhou2015filtering}.   
Based on this trend, we counted the number of formal English words in each of the content fields in a given post, and the article the post pointed to.
In order to do so, we used the PyDictionary~\cite{pydictionary}, which is a a dictionary module for Python, which provides meanings, translations, synonyms and antonyms of words.
It uses WordNet~\cite{wordnet} for definitions, Google~\cite{google_translate} for translations, and thesaurus.com~\cite{thesaurus} for synonyms and antonyms. 

The following function counts the number of formal English words using the extraction of the words in each content element and searching each word in the English dictionary. 

\[
number\text{-}of\text{-}formal\text{-}words(x) = lang\text{-}dict_{formal}(words(x))
\]

In a similar way, we used the following function in order to count the number of informal English words.
The difference between this current function and the previous is that here we count the number of words that are not exist in the English dictionary.

\[
number\text{-}of\text{-}informal\text{-}words(x) = lang\text{-}dict_{informal}(words(x))
\]

\paragraph{Formal, and Informal English Word Ratio} 
\label{sec:formal_and_informal_word_ratio}
An additional optional function can be used is the ratio of the formal and informal English words among all the words in a given content.
The following function measures the ratio of formal words among all the words that compose the entire content. 
 
\[
percent\text{-}of\text{-}formal\text{-}words(x) = \frac{lang\text{-}dict_{formal}(words(x))}{words(x)}
\]

Similarly, the following function calculates the ratio of informal English words among all the words that compose the entire content.

\[
percent\text{-}of\text{-}informal\text{-}words(x) = \frac{lang\text{-}dict_{informal}(words(x))}{words(x)}
\]

\subsubsection{Features Taken from Abuser Detection}
\label{sec:features_taken_from_abuser_detection}
Clickbait is responsible of the rapid spread of rumors and misinformation online~\cite{chen2015misleading}. 
This malicious activity is common also among \emph{abusers} in OSM~\cite{elyashar2017measurement}. 
Therefore, we can use features that are helpful for \emph{abuser} detection also for the clickbait detection~\cite{potthast2016clickbait}.



\paragraph{Post Creation Hour} 
\label{sec:post_creation_hour}

In many cases, crowdturfing campaigns are carried out in a short interval of time (ranging from a period of hours to a single week)~\cite{wang2012serf}.
This means that we should be able to observe a high volume of activities associated with a given crowdturfing campaign at clear times. 

\paragraph{Post Longevity} 
\label{sec:post_longevity}
A common feature for detecting \emph{abusers} is their life-span.
Simple bots are known to have a shorter life-span than \emph{legitimate users}~\cite{lee2013crowdturfers}.  
In this study, we do not have accounts, but posts.
Therefore, we calculate the age of each post as a feature.

\paragraph{Activity Based Characteristics / Behavior Analysis} 
\label{sec:activity_based_characteristics}
Another method for detecting \emph{abusers} in OSM is by extracting features that are based on their activity.
For example, in Twitter, the average number of links per tweet, average number of user mentions per tweet, etc.~\cite{lee2013crowdturfers}.
In this study, we extracted the following features: 

1) Number of @ Signs - this sign is used to call out user names in posts in Twitter~\cite{twitter_glossary}.

2) Number of Hashtags - a hashtag is any word or phrase with prefix consisting of the \# symbol. 
When a Twitter account clicks on a hashtag, Twitter will present to him or her other tweets containing the same keyword or topic~\cite{twitter_glossary}.

3) Number of Retweets - a retweet is defined as a tweet that a user forwards to his or her followers. 
Retweets are often used to pass along news, or other valuable information on Twitter~\cite{twitter_glossary}.  

4) Number of Question Marks, Commas, Colons, and Ellipses - one of the methods used to attract readers' attention to a specific article is the use of question marks, commas, colons and ellipses in titles.

We extracted the features mentioned previously from all the content field available in the given posts, and articles.
For example, we counted the number of @ signs in a post's title, text extracted from an image, an article's title, etc. 

\paragraph{Article Properties}
\label{sec:article_properties}

Additional features that were reported to perform well in the past are those features that are related to the account properties (e.g., number of friends, number of followers)~\cite{lee2013crowdturfers}.
In this study, we can look on the targeted article as the account in OSM.
In this current study, there is information regarding the post, as well as the targeted article.
Thus, we extracted the following features: the \emph{number of article keywords}, the \emph{number of paragraphs}, and the \emph{number of article captions}. 

\section{Evaluation Results}
\label{sec:evaluation_results}

The evaluation process was carried out in TIRA~\cite{potthast:2014} as provided by the challenge organizers. 
In order to evaluate the predictive power of the extracted features, we applied information gain feature selection~\cite{stachniss2005information}. 
The most significant features are presented in Table~\ref{table:info_gain_results}.

\begin{table}[htb]
	
	\centering
	\caption{TOP FEATURES ORDERED BY INFO GAIN.}
	\scalebox{0.78}{
	\begin{tabular}{ |m{0.7cm}|m{0.4cm}|m{5.5cm}|m{0.7cm}| }
		\hline
		\textbf{Dataset} & \textbf{Rank} & \textbf{Feature Name} & \textbf{Info Gain} \\ \hline
		\parbox[t]{2mm}{\multirow{12}{*}{\rotatebox[origin=c]{90}{Training}}} & 1 & diff num of characters post title \& article keywords & 0.036 \\
								  & 2 & num of characters ratio post image text \& post title & 0.032  \\
								  & 3 & num of characters in post title & 0.030 \\
								  & 4 & num of question marks in post title & 0.211 \\
								  & 5 & diff num of characters post title \& post image text  & 0.02 \\
								  & 6 & num of characters ratio article description \& post title & 0.019 \\
								  & 7 & num of characters ratio article paragraphs \& post title & 0.018 \\ 
								  & 8 & diff num of words post title \& article keywords & 0.018  \\ 
								  & 9 & num of words ratio article description \& post title & 0.017 \\ 
								  & 10 & num of characters ratio article paragraphs \& article desc  & 0.017\\ 
							      & 11 & num of words in post title & 0.017\\
							      & 12 & num of formal words in post title & 0.016\\ \hline
		\parbox[t]{2mm}{\multirow{12}{*}{\rotatebox[origin=c]{90}{Validation}}}  & 1 & num of characters in post title & 0.082 \\
									  & 2 & num of characters ratio post image text \& post title & 0.082  \\
									  & 3 & diff num of characters post title \& article keywords & 0.065 \\
									  & 4 & diff num of characters post title \& post image text & 0.061 \\
									  & 5 & num of words ratio post image text \& post title & 0.058 \\
									  & 6 & num of words in post title  & 0.056 \\
									  & 7 & num of formal words in post title & 0.056 \\
									  & 8 & num of words ratio article description \& post title & 0.055 \\ 
									  & 9 & num of characters ratio article description \& post title & 0.055  \\ 
									  & 10 & num of characters ratio article title \& post title & 0.052 \\ 
									  & 11 & num of words ratio article title \& post title & 0.048\\ 
									  & 12 & diff num of words post title \& article keywords  & 0.047\\ \hline
	\end{tabular}
	}
	\label{table:info_gain_results}
	\end{table}

As can be seen, the most significant features in both datasets were: the \emph{difference number of characters between post title and article keywords} (first place in the training dataset and third place in the validation dataset), the \emph{number of characters in post title} (third place in the training dataset and first place in the validation dataset), the \emph{number of characters ratio post image text and post title} (second place in both datasets), the \emph{difference number of characters between post title and post image text} (fifth place in the training dataset and fourth place in the validation dataset), and the \emph{number of characters ratio between article's description and post's title} (sixth place in the training dataset and eighth place in the validation dataset).
In addition, several features were found to be more significant in one dataset than the other. 
For example, number of question marks in the post's title is the fourth most significant feature in the training dataset, whereas it is only the the 71-th most significant feature in the validation dataset.

We trained several ML classifiers in order to determine the differences between \emph{clickbait posts} and \emph{legitimate posts} (XGBoost showed the best results). 
Each classifier was trained with multiple sets of features having the highest information gain score. 
The performance of the classifiers was evaluated in terms of the area under ROC curve (AUC), accuracy, precision, and recall during internal 10-fold cross-validation. 
The results of the best classifier for each algorithm are summarized in Table~\ref{table:classifier_performances}.
We note that the best classifier in both datasets was trained using XGBoost on all of the features with the highest information gain score. 
The best performance of the XGBoost classifier for the training dataset is as follows: an AUC of 0.715, accuracy of 0.732, and precision and recall of 0.75, and 0.92, while the best performance of this classifier for the validation dataset is: an AUC of 0.8, accuracy of 0.812, and precision and recall of 0.819, and 0.966.

\begin{table}[htb]
	\centering
	\caption{PERFORMANCE OF THE BEST CLASSIFIERS.}
	\scalebox{0.65}{
		\begin{tabular}{ |c|c|c|c|c|c|c| }
			\hline
			\textbf{Dataset} & \textbf{Algorithm} & \textbf{\#Features} & \textbf{AUC} & \textbf{Accuracy} & \textbf{Precision} & \textbf{Recall}\\ \hline
			\multirow{6}{*}{Training} & XGBoost & All & 0.715 & 0.732 & 0.75 & 0.92 \\
									  & XGBoost & 20 & 0.707 & 0.728 & 0.744 & 0.925  \\
									  & Random Forest & All & 0.707 & 0.732 & 0.752 & 0.913  \\
									  & AdaBoost & 20 & 0.702 & 0.721 & 0.74 & 0.917  \\
							          & Random Forest & 20 & 0.698 & 0.725 & 0.753 & 0.895  \\ 
									  & Decision Tree & All & 0.583 & 0.636 & 0.743 & 0.721  \\ \hline
			\multirow{6}{*}{Validation} & XGBoost & All & 0.8 & 0.812 & 0.819 & 0.966 \\
										& Random Forest & All & 0.789 & 0.811 & 0.823 & 0.955  \\
										& AdaBoost & All & 0.777 & 0.802 & 0.818 & 0.951  \\
										& XGBoost & 20 & 0.776 & 0.807 & 0.814 & 0.966  \\
										& Random Forest & 20 & 0.771 & 0.804 & 0.823 & 0.944  \\ 
										& Decision Tree & All & 0.635 & 0.725 & 0.824 & 0.811  \\ \hline
									
		\end{tabular}
	}
	\label{table:classifier_performances}
\end{table}

In addition, we measured the length of the post's title in each datasets. 
The average lengths are presented in Table~\ref{table:post_title_length_measure}.

\begin{table}[htb]
	\centering
	\caption{POST'S TITLE LENGTH MEASURES.}
	\scalebox{0.65}{
		\begin{tabular}{ |c|c|c|c| }
			\hline
			\textbf{Dataset} & \textbf{Feature Name} & \textbf{Clickbait} & \textbf{Non-clickbait} \\ \hline
			\multirow{2}{*}{Training} & num of chars in post title & 71.83 & 81.746 \\
									  & num of words in post title & 11.787 & 12.877 \\ \hline
			\multirow{2}{*}{Validation} & num of chars in post title & 59.288 & 74.69 \\
										& num of words in post title & 10.012 & 11.898 \\\hline
			
		\end{tabular}
	}
	\label{table:post_title_length_measure}
\end{table}

Surprisingly, as opposed to previous studies~\cite{chakraborty2016stop,horne2017just} which suggested that \emph{legitimate content} is shorter than \emph{malicious content}, we found the opposite. 
The average number of characters in the titles of \emph{clickbait's post} is 71.83 and 59.288 respectively for the training and validation datasets, whereas the average number of characters in \emph{non-clickbait post}'s title is 81.746 and 74.69 in the training and validation datasets respectively.
Similarly, the average number of words in the titles of clickbait post is 11.787 and 10.012, whereas the average number of words in \emph{non-clickbait post}'s titles is 12.877 and 11.898 respectively in the training and validation datasets.
The differences of the word length, as well as the character length in both datasets were found to be statistically significant. 

\section{Conclusion}
\label{sec:conclusion}
In this paper, we attempt to detect \emph{clickbait posts} as part of the Clickbait Challenge 2017.
We proposed a variety of useful features in order to distinguish between \emph{clickbait} and \emph{legitimate posts} published in OSM.
Based on the evaluation phase, which is presented in Section~\ref{sec:evaluation_results}, we conclude the following:

First, as opposed to previous studies~\cite{chakraborty2016stop,horne2017just}, we found that \emph{clickbait posts} are shorter than \emph{legitimate posts} (see Table~\ref{table:post_title_length_measure}).
The differences between the lengths in both datasets were found statistically significant.
We think that in the future it would be interesting to measure these features again in other up-to-date datasets in order to understand whether this generally reflects a new phenomenon, or whether this finding is unique to the two datasets provided.   
Additional option to explain these results can be associated to the domain of academic papers. 
Letchford et al. found that academic papers with shorter titles~\cite{letchford2016advantage}, or simpler language in their abstracts~\cite{letchford2015advantage} get cited more often.  

Second, the features that were found to be the most significant and useful for clickbait detection are the \emph{difference number of characters between post's title and article keywords}, the \emph{number of characters ratio between text extracted from post's image and post's title}, and the \emph{number of characters in post's title}.
The top feature (\emph{difference number of characters between post's title and article keywords}) suggests a connection between the post and the article the post points.
These features, as well as other similar  features, suggest that there is a subtlety that can be measured by analyzing both the post and the article, which can be helpful for detecting clickbait.
The second best feature (\emph{number of characters ratio between text extracted from post's image and post's title}) suggests that there is a connection between a given post and the image attached. 
Furthermore, we found that extracting the text present in the attached image can be useful for detecting clickbait.
In addition, the next best feature (\emph{number of characters in post's title}) reinforces our view that despite the contradiction we found regarding the length of the \emph{clickbait posts}, features that measure the length of the post's title is useful for clickbait detection (e.g., \emph{number of characters}, and \emph{number of words}).
These two features were found to be significant according to the information gain score in both datasets (see Table~\ref{table:info_gain_results}).    

Third, our suggested feature, the \emph{number of formal words in post's title}, was found to be significant according to information gain score (place 12, and 6 in the training, and validation datasets respectively).
We can conclude that in addition to extract part of speech (POS), it is recommended to calculate how many formal English words present in the post's title.   

Finally, we found that the post's title is the most important component to use for detecting clickbait.
It sounds trivial, but we succeeded at showing this empirically.  
We created the same features for each competent equally.
This means that ,for example, the \emph{number of words} feature was calculated for the post's title, as well as the article's title, description, etc.
Of the twenty-four most influential features (based on the information gain score for both datasets) only one feature is not related to post's title (see Table~\ref{table:info_gain_results}) .

\section*{Acknowledgment}
The authors would like to thank Robin Levy-Stevenson for proofreading this paper.

\begin{raggedright}
\bibliography{clickbait17-notebook-lit}
\end{raggedright}
\end{document}